\documentclass[twocolumn,showpacs,aps,prl,superscriptaddress]{revtex4}

\usepackage{graphicx}
\usepackage{amsmath}

\begin{document}

\title{Spin Polarization Dependence of the Coulomb Drag at Large $r_{s}$}

\author{R. Pillarisetty}
\affiliation{Department of Electrical Engineering, Princeton
University, Princeton, New Jersey 08544}
\author{Hwayong Noh}
\affiliation{Department of Electrical Engineering, Princeton
University, Princeton, New Jersey 08544}
\affiliation{Department
of Physics, Sejong University, Seoul 143-747, Korea}
\author{E. Tutuc}
\affiliation{Department of Electrical Engineering, Princeton
University, Princeton, New Jersey 08544}
\author{E.P. De Poortere}
\affiliation{Department of Electrical Engineering, Princeton
University, Princeton, New Jersey 08544}
\author{D.C. Tsui}
\affiliation{Department of Electrical Engineering, Princeton
University, Princeton, New Jersey 08544}
\author{M. Shayegan}
\affiliation{Department of Electrical Engineering, Princeton
University, Princeton, New Jersey 08544}

\date{\today}

\begin{abstract}
We find that the temperature dependence of the drag resistivity
($\rho_{D}$) between two dilute two-dimensional hole systems
exhibits an unusual dependence upon spin polarization. Near the
apparent metal-insulator transition, the temperature dependence of
the drag, given by $T^{\alpha}$, weakens with the application of a
parallel magnetic field ($B_{||}$), with $\alpha$ saturating at
half its zero field value for $B_{||} > B^{*}$, where $B^{*}$ is
the polarization field. Furthermore, we find that $\alpha$ is
roughly 2 at the parallel field induced metal-insulator
transition, and that the temperature dependence of
$\rho_{D}/T^{2}$ at different $B_{||}$ looks strikingly similar to
that found in the single layer resistivity. In contrast, at higher
densities, far from the zero field transition, the temperature
dependence of the drag is roughly independent of spin
polarization, with $\alpha$ remaining close to 2, as expected from
a simple Fermi liquid picture.
\end{abstract}
 \maketitle

The spin degree of freedom of the electron plays a fundamental
role in condensed matter physics, and significantly influences the
electronic properties of solids. For simple metals, which are well
described by Fermi liquid theory, the role spin plays in electron
transport is reasonably well understood. In contrast, the spin
degree of freedom plays a unique role in stabilizing several
exotic non-Fermi liquid states, such as BCS
superconductivity\cite{bcs}. Recently, much attention has focused
upon dilute two-dimensional (2D) systems and the possibility that
the spin degree of freedom stabilizes a novel phase of matter.
These systems, which have large ratios of carrier interaction
energy to kinetic energy ($r_{s} > 10$), exhibit an anomalous
metallic behavior and an apparent metal-insulator transition
(MIT)\cite{mit}, contradictory to the scaling theory of
localization\cite{scaling}. Despite numerous studies, no
conclusive understanding of the behavior in this regime exists.
However, many believe that the spin degree of freedom plays an
important role in stabilizing the metallic behavior. This results
from numerous observations\cite{mit,simonian,yoon} that the
metalliclike behavior is suppressed as the 2D system is spin
polarized by the application of a parallel magnetic field
($B_{||}$). Furthermore, recent experiments, in a variety of
different systems\cite{okamoto,vitkalov,shashkin,zhu,vakili}, show
a strong enhancement of the spin susceptibility as the carrier
density is reduced near the MIT, suggesting a possible transition
to a ferromagnetic ground state. To help elucidate the bizarre
role spin plays in dilute 2D systems, we have studied the spin
polarization dependence of the Coulomb drag near the 2D MIT.

Drag resistivity measurements \cite{TJ} between parallel 2D layers
provide a powerful probe of carrier-carrier interactions. These
experiments are performed by driving a current ($I$) in one layer,
and measuring the potential ($V_{D}$), which arises in the other
layer due to momentum transfer. The drag resistivity ($\rho_{D}$),
given by $V_{D}/I$, is directly proportional to the interlayer
carrier-carrier scattering rate. In a Fermi liquid picture, the
drag should follow a $T^{2}$ dependence at low
temperatures\cite{boltzmann}. Furthermore, the Fermi liquid
framework implies that the temperature dependence of the drag
should remain independent of spin polarization. While well
understood exceptions to the $T^{2}$ dependence
exist\cite{phononplasmon}, measurements of the drag at $B_{||}=$
0, near the apparent 2D MIT, exhibit an anomalous $T$ dependence,
which is greater than $T^{2}$ and correlated with the metalliclike
$T$ dependence in the single layer resistivity\cite{ravi}.

In this article, we study the effect of spin polarization on the
temperature dependence of the drag between dilute 2D hole systems.
Near the apparent MIT, we find that the temperature dependence of
the drag, given by $T^{\alpha}$, weakens significantly with the
application of $B_{||}$, with $\alpha$ saturating at half its zero
field value above the polarization field ($B^{*}$). Furthermore,
we find that $\alpha$ is roughly 2 at the parallel field induced
MIT, and that the $T$ dependence of $\rho_{D}/T^{2}$ at different
$B_{||}$ looks strikingly similar to that found in the single
layer resistivity. In contrast, at higher densities, far from the
zero field MIT, the $T$ dependence of the drag is roughly
independent of spin polarization, with $\alpha$ remaining close to
2, as expected from a simple Fermi liquid picture.

Two samples were used in this study. Sample A (B) contains a
double quantum well structure, consisting of two 150 (175) \AA\ Si
doped p-type GaAs quantum wells separated by a pure 150 (100) \AA\
AlAs barrier, which was grown by molecular beam epitaxy on a
(311)A GaAs substrate. Sample A (B) has an average grown layer
density and center to center layer separation of 2.5
(7.0)$\times10^{10}$ cm$^{-2}$ and 300 (275) \AA, respectively.
The average mobilities in each layer of Sample A (B) at 300 mK,
are 1.5 (6.7)$\times10^{5}$ cm$^{2}$/Vs. Single layer transport in
Sample A (see \cite{ravi}) shows a clear zero field MIT at $p=
8.5\times10^{9}$ cm$^{-2}$. The samples were processed allowing
independent contact to each of the two layers, using a selective
depletion scheme\cite{ic}. In addition, both layer densities are
independently tunable using evaporated metallic gates.

The data presented in this paper were obtained in top loading
dilution and $^{3}$He refrigerators. The densities in each layer
were determined by independently measuring Shubnikov-de Haas
oscillations. Drive currents between 50 pA to 10 nA were passed,
in the [$\bar{2}$33] direction, through one of the layers, while
the drag signal was measured in the other layer, using lock-in
techniques. To ensure no spurious sources contributed to our
signal, all the standard consistency checks associated with the
drag technique were performed\cite{TJ}.

\begin{figure}[!t]
\begin{center}
\includegraphics[width=3.4in,trim=0.2in 0.2in 0.2in 0.2in]{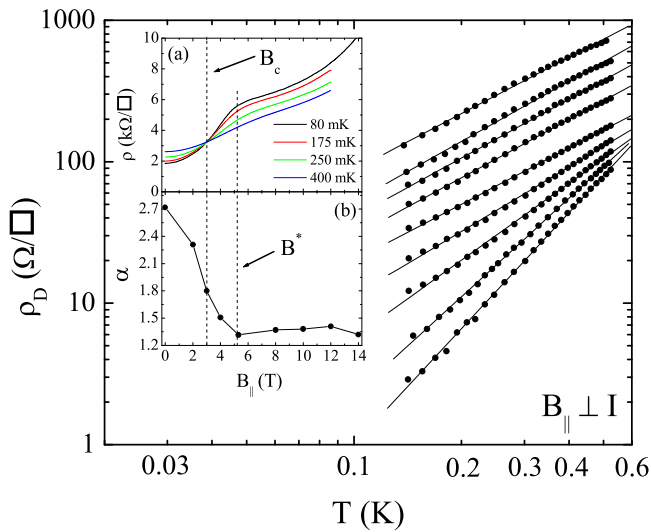}
\end{center}
\caption{\label{1}$\rho_{D}$ vs $T$ on log-log scale at $p_{m}=
2.15\times10^{10}$ cm$^{-2}$, with $B_{||} \perp I$. Traces are
for (from bottom) $B_{||}=$ 0, 2, 3, 4, 5.3, 8, 10, 12, and 14 T.
Solid lines are the best linear fits of each trace. Inset (a):
$\rho$ vs $B_{||}$ for different $T$. Inset (b): $\alpha$ vs
$B_{||}$. $\alpha$ deduced from the slope of the linear fits in
the main plot. $B_{c}=$ 3 T and $B^{*}=$ 5.3 T indicated by the
dashed lines.}
\end{figure}

We begin our presentation of the data by first looking at the
temperature dependence of the drag at different $B_{||}$. This is
presented in Fig 1 where we plot $\rho_{D}$ vs $T$ on log-log
scale, for matched layer densities ($p_{m}$) of
$2.15\times10^{10}$ cm$^{-2}$. Here $B_{||}$ is aligned
perpendicular to the drive current. At this density, the single
layer resistivity exhibits a strong metalliclike $T$ dependence at
zero field, and earlier studies show that $\rho_{D}$ exhibits an
anomalously large enhancement with $B_{||}$\cite{ravi2}. Single
layer magnetotransport, shown in inset (a), exhibits a clear
$B_{||}$ induced MIT at $B_{c}=$ 3 T, and the onset of full spin
polarization at $B^{*}=$ 5.3 T. The data in the main plot clearly
shows that the $T$ dependence of the drag has an unusual
dependence on $B_{||}$. At low fields, the slopes of the linear
fits of the data weaken significantly as $B_{||}$ is increased. In
contrast, at higher $B_{||}$, the slopes of these fits seem to be
independent of increasing $B_{||}$. To examine this more
carefully, we plot the slopes of each of these fits against
$B_{||}$ in inset (b). Here the slope corresponds to the exponent,
$\alpha$, where $\rho_{D} \propto T^{\alpha}$. At zero field,
$\alpha$ is significantly larger than 2, as has been discussed
earlier\cite{ravi}. From this plot it is clear that as $B_{||}$ is
increased, $\alpha$ weakens significantly, and then saturates for
$B_{||} > B^{*}$. In addition to this observation, there are two
other very interesting features to this plot. First, the value of
$\alpha$ for $B_{||} > B^{*}$ is roughly half that found at zero
field. Secondly, $\alpha$ is close to 2 (about 1.8) when $B_{||}=
B_{c}$. To show this more explicitly, in Fig 2 we plot
$\rho_{D}/T^{2}$ vs $T$, for $B_{||}=$ 0, 2, 3, 4, and 5.3 T. The
first point we make is that this data looks strikingly similar to
a plot of the $T$ dependence of the single layer resistivity at
different $B_{||}$. Furthermore, $\rho_{D}/T^{2}$ changes from
increasing with $T$ to decreasing with $T$ very close to $B_{c}=$
3 T, where the parallel field induced MIT is observed in the
single layer resistivity.

\begin{figure}[!t]
\begin{center}
\includegraphics[width=2.6in,trim=0.2in 0.2in 0.2in 0.2in]{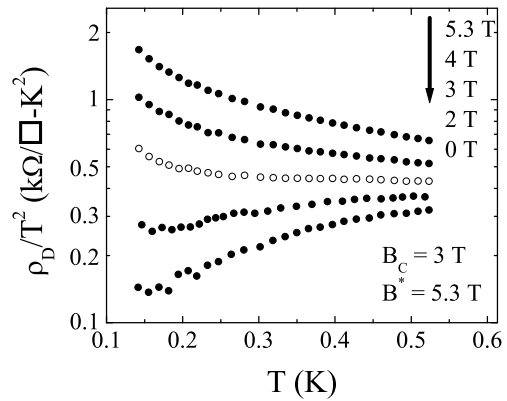}
\end{center}
\caption{\label{2}$\rho_{D}/T^{2}$ vs $T$ for (from bottom)
$B_{||}=$ 0, 2, 3, 4, and 5.3 T. $p_{m}= 2.15\times10^{10}$
cm$^{-2}$, with $B_{||} \perp I$. Data at $B_{||}= B_{c}$ shown as
open circles.}
\end{figure}

Although it surely appears that these observations, primarily the
saturation of the $T$ dependence above $B^{*}$, are due to a spin
effect, one must be careful to consider orbital
effects\cite{dassarma}. Due to the finite layer thickness of our
2D systems, the coupling of $B_{||}$ to the orbital motion of the
carriers, can significantly effect the in-plane magnetotransport.
The high field magnetoresistance observed for $B_{||} > B^{*}$ in
2D GaAs samples\cite{yoon,zhu}, is typically believed to arise
from such an orbital effect. By reorienting $B_{||}$ such that it
is parallel to the drive current, the contribution of orbital
effects to the in-plane magnetoresistance can be significantly
reduced, as compared to when $B_{||} \perp I$\cite{dassarma}. This
is shown in inset (a) of Fig 3, where we plot the single layer
magnetoresistance for $p_{m}= 1.85\times10^{10}$ cm$^{-2}$ with
$B_{||} \parallel I$, at different $T$. It is clear that the high
field magnetoresistance arising from orbital effects is much
weaker in the $B_{||} \parallel I$ configuration. The data at
$p_{m}= 1.85\times10^{10}$ cm$^{-2}$ with $B_{||}
\parallel I$ also shows a clear $B_{||}$ induced MIT at $B_{c}=$ 1.7 T, along with a clear shoulder in the
magnetoresistance, corresponding to full spin polarization, at
$B^{*}=$ 3.4 T. In the main plot of Fig 3, we present the $T$
dependence of $\rho_{D}$ at different $B_{||}$ for $p_{m}=
1.85\times10^{10}$ cm$^{-2}$ with $B_{||}
\parallel I$. Note here that the exact same behavior as seen
with $B_{||} \perp I$ is observed, with the strength of the $T$
dependence weakening as $B_{||}$ is increased and then saturating
above $B^{*}$. To show this more explicitly, we plot $\alpha$ vs
$B_{||}$ in inset (b). Again, $\alpha$ is found to saturate at
roughly half its zero field value for $B_{||} > B^{*}$. In
addition, $\alpha$ becomes roughly 2 (again about 1.8) at $B_{||}=
B_{c}$. This data unambiguously shows that orbital effects are not
playing a role here, and that the unusual behavior we observe
arises from a spin effect.

\begin{figure}[!t]
\begin{center}
\includegraphics[width=3.4in,trim=0.2in 0.2in 0.2in 0.2in]{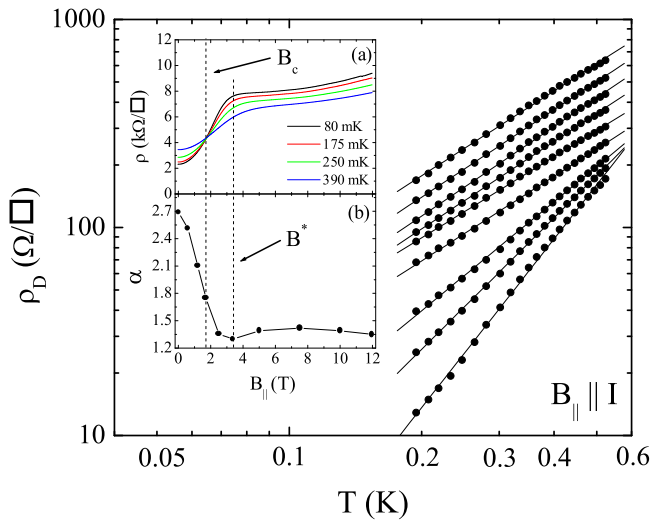}
\end{center}
\caption{\label{3}$\rho_{D}$ vs $T$ on log-log scale at $p_{m}=
1.85\times10^{10}$ cm$^{-2}$ with $B_{||} \parallel I$. Traces are
for (from bottom) $B_{||}=$ 0, 1.2, 1.7, 2.5, 3.4, 5, 7.5, 10, and
12 T. Solid lines are the best linear fit of each trace. Inset
(a): $\rho$ vs $B_{||}$ for different $T$. Inset (b): $\alpha$ vs
$B_{||}$. $B_{c}=$ 1.7 T and $B^{*}=$ 3.4 T indicated by the
dashed lines.}
\end{figure}

To demonstrate the universality of this effect at low densities,
near the apparent MIT, we plot all of our $T$ dependence data
taken for $p_{m}=$ 2.15, 1.75, 1.5, and 1.25$\times10^{10}$
cm$^{-2}$ with $B_{||} \perp I$ and for $p_{m}= 1.85\times10^{10}$
cm$^{-2}$ with $B_{||} \parallel I$ on a scaled plot, which is
presented in Fig 4. Here we plot $\alpha$ normalized by its zero
field value against $B_{||}/B^{*}$ for each data set. These data
sets range from $r_{s}=$ 9.2 to 12.1, using $m^{*}= 0.17m_{e}$. As
is shown in the plot, it is clear that all five of these data sets
collapse onto each other. In addition, for all the data it is
clear that the $\alpha$ found at zero field decreases by a factor
of roughly 2, at the polarization field, and saturates upon
further increase of $B_{||}$.

\begin{figure}[!t]
\begin{center}
\includegraphics[width=2.6in,trim=0.2in 0.2in 0.2in 0.2in]{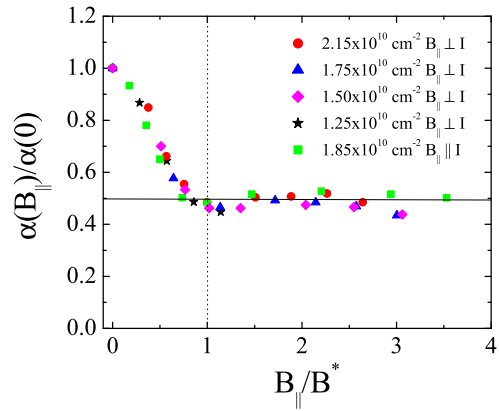}
\end{center}
\caption{\label{4}$\alpha(B_{||})/\alpha(0)$ vs $B_{||}/B^{*}$ for
$p_{m}=$ 2.15, 1.75, 1.5, and 1.25$\times10^{10}$ cm$^{-2}$, with
$B_{||} \perp I$ and for $p_{m}= 1.85\times10^{10}$ cm$^{-2}$,
with $B_{||} \parallel I$.}
\end{figure}

It is not hard to imagine that this behavior we observe must be
related to the large $r_{s}$ value of our dilute 2D system. In the
small $r_{s}$ limit, where carrier-carrier interactions are weak,
Fermi liquid theory dictates that the $T$ dependence of the drag
should be independent of spin polarization. To investigate whether
the bizarre behavior found in the dilute regime is suppressed when
we go to lower $r_{s}$, we have investigated a sample with $p_{m}=
7.0\times10^{10}$ cm$^{-2}$ ($r_{s}= 5.1$), which is far from the
zero field MIT. At this density, the zero field properties of the
drag, including a nearly $T^{2}$ dependence, are reasonably well
described by Boltzmann theory\cite{boltzmann}. In inset (a) of Fig
5, we present the single layer in-plane magnetotransport in this
sample, for different $T$. These measurements were done with
$B_{||} \parallel I$, at $T=$ 0.3, 0.6, 0.9, and 1.2 K. A clear
crossing point is found at $B_{||}=$ 11.6 T. In addition, this
data clearly shows the start of the shoulder feature at the
highest fields. Our estimates place $B^{*}$ somewhere between 17.5
and 18 T, which would correspond to roughly 90 \% polarization at
$B_{||}=$ 16 T. In the main plot of Fig 5, we present the $T$
dependence of $\rho_{D}$ at $p_{m}= 7.0\times10^{10}$ cm$^{-2}$,
for $B_{||}=$ 0, 10 and 16 T. It is clear that, to first order,
there is little change to the nearly $T^{2}$ dependence found at
zero field. The $\alpha$'s deduced from the linear fits of these
data, which are shown in inset (b), give $\alpha=$ 1.9, 1.9, and
1.7 at $B_{||}=$ 0, 10, and 16 T, respectively. It appears that
whatever produces the unusual effect observed at low densities
plays a perturbative role here. Here $\alpha$ decreases by roughly
a factor of 1.1 as $B_{||}$ approaches $B^{*}$, in contrast to the
factor of 2 decrease found at lower densities close to the MIT.

\begin{figure}[!t]
\begin{center}
\includegraphics[width=3.4in,trim=0.2in 0.2in 0.2in 0.2in]{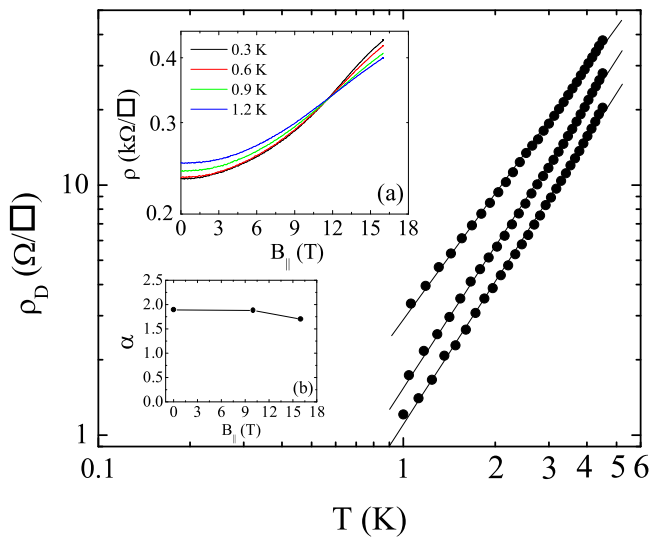}
\end{center}
\caption{\label{5}$\rho_{D}$ vs $T$ on log-log scale, for $p_{m}=
7.0\times10^{10}$ cm$^{-2}$, with $B_{||} \parallel I$. Traces are
for (from bottom) $B_{||}=$ 0, 10, and 16 T. Solid lines are the
best linear fit of each trace. Inset (a): $\rho$ vs $B_{||}$, for
$T=$ 0.3, 0.6, 0.9, and 1.2 K. Inset (b): $\alpha$ vs $B_{||}$,
deduced from the linear fits in the main plot.}
\end{figure}

We would now like to discuss these experimental results and the
consequences they yield, with regards to the 2D MIT problem. The
first piece of data we comment on is the striking similarity of
the $T$ dependence of $\rho_{D}/T^{2}$ at different $B_{||}$ with
respect to that in the single layer resistivity, which is shown in
Fig 2. In a Fermi liquid picture, one expects the drag to exhibit
a $T^{2}$ dependence and the resistivity to remain relatively
temperature independent. However, at zero field the $T$ dependence
of both $\rho$ and $\rho_{D}$ are anomalously enhanced with
respect to what should be expected for both properties. Our data
clearly show that the $T$ dependence of both $\rho$ and $\rho_{D}$
exhibit exactly the same qualitative change as the 2D system is
spin polarized, which is surprising since these are extremely
different properties. This implies that there is some fundamental
property in the system, which has a strong spin polarization
dependence, which affects $\rho$ and $\rho_{D}$ in exactly the
same way.

In general, the fact that the $T$ dependence of the drag exhibits
such a strong dependence on spin polarization, with $\alpha$
saturating at half its zero field value above the polarization
field, seems to be a significant departure from Fermi liquid
behavior. While we currently have no clear understanding of this
bizarre behavior found in the dilute regime, we would like to
comment on one possible explanation. Recently, a theory has been
put forth, which states that the zero field MIT is a transition
from a Fermi liquid to a Wigner crystal, via a series of
intermediate phases\cite{spivak}. In other words, near the
transition, there is a phase coexistence of the Fermi liquid and
Wigner solid. Due to the much larger spin entropy and spin
susceptibility of the solid phase, free energy arguments dictate
that the fraction of Wigner crystal in the 2D system ($f_{WC}$)
will grow significantly with increasing temperature and spin
polarization. This behavior is analogous to the Pomeranchuk effect
in $^{3}$He. Furthermore, the $T$ dependence of $f_{WC}$ weakens
significantly as $B_{||}$ is applied, due to a reduction in the
spin entropy of the system. For $B_{||} > B^{*}$, this $T$
dependence saturates, since the spin entropy of the system is
lost. This theory\cite{spivak} states that the Wigner solid
regions of the 2D system dominate the resitivity, and the
underlying $T$ and $B_{||}$ dependences of $f_{WC}$ yield the
metalliclike transport anomalies, which have been observed in
numerous experiments\cite{mit}. An extension of this
theory\cite{unpublished} shows that these Wigner crystal regions
dominate the drag resistivity as well, and that the anomalies to
the $T$ and $B_{||}$ dependences of $\rho_{D}$ similarly arise
from the underlying $T$ and $B_{||}$ dependences of $f_{WC}$. This
could possibly explain why the $T$ dependence of $\rho$ and
$\rho_{D}/T^{2}$ at different $B_{||}$ look so similar.
Furthermore, our observation of a strong decrease in the strength
of the $T$ dependence of $\rho_{D}$ as $B_{||}$ is increased, and
its saturation for $B_{||} > B^{*}$ seem to directly follow from
the fact that the $T$ dependence of $f_{WC}$ weakens as $B_{||}$
is applied, and saturates for $B_{||} > B^{*}$.

In conclusion, we have investigated the effect of spin
polarization on the $T$ dependence of the Coulomb drag near the
MIT. We find that the temperature dependence of the drag, given by
$T^{\alpha}$, weakens as $B_{||}$ is applied, with $\alpha$
saturating at half its zero field value for $B_{||} > B^{*}$. In
addition, the $T$ dependence of $\rho_{D}$ becomes roughly $T^{2}$
at the parallel field induced MIT. We find this effect to be
independent of field orientation, ruling out the possibility that
it arises from an orbital effect. Finally, at higher densities,
far from the MIT, the $T$ dependence of $\rho_{D}$ is roughly
independent of spin polarization, as expected from a simple Fermi
liquid picture.

We are grateful to B. Spivak and S.A. Kivelson for discussions and
for supplying us with a copy of their unpublished manuscript. In
addition, we thank K. Vakili for assistance. This research was
funded by the NSF, a DURINT grant from the ONR, and the DOE.

\end{document}